\documentclass{article}
\usepackage{ijcai21}

\usepackage{times}

\usepackage{soul}
\usepackage{url}
\usepackage[hidelinks]{hyperref}
\usepackage[utf8]{inputenc}
\usepackage[small]{caption}
\usepackage{booktabs}
\usepackage{multirow}
\usepackage{amsmath,amssymb,amsfonts}
\usepackage[linesnumbered, ruled]{algorithm2e}
\usepackage{graphicx}
\usepackage{subcaption}
\usepackage{kantlipsum}
\SetKwInput{KwInput}{Input}                
\SetKwInput{KwOutput}{Output}
\title{Segmented Federated Learning for Adaptive Intrusion Detection System}

\author{
Geet Shingi$^1$
 Harsh Saglani$^1$\and
 Preeti Jain$^1$
\affiliations
$^1$Dept. of Computer Engineering, \\
Pune Institute of Computer Technology, \\
Maharashtra, India.
\emails
geet.shingi@gmail.com, harsh.saglani6@gmail.com, pajain@pict.edu
}

\begin{document}

\maketitle

\begin{abstract}
 Cyberattacks are a major issues and it causes organizations great financial, and reputation harm. However, due to various factors, the current network intrusion detection systems (NIDS) seem to be insufficent. Predominant NIDS identifies Cyberattacks through a handcrafted dataset of rules. Although the recent applications of machine learning and deep learning have alleviated the enormous effort in NIDS, the security of network data has always been a prime concern. However, to encounter the security problem and enable sharing among organizations, Federated Learning (FL) scheme is employed. Although the current FL systems have been successful, a network's data distribution does not always fit into a single global model as in FL. Thus, in such cases, having a single global model in FL is no feasible. In this paper, we propose a Segmented-Federated Learning (Segmented-FL) learning scheme for a more efficient NIDS. The Segmented-FL approach employs periodic local model evaluation based on which the segmentation occurs. We aim to bring similar network environments to the same group. Further, the Segmented-FL system is coupled with a weighted aggregation of local model parameters based on the number of data samples a worker possesses to further augment the performance. The improved performance by our system as compared to the FL and centralized systems on standard dataset further validates our system and makes a strong case for extending our technique across various tasks. The solution finds its application in organizations that want to collaboratively learn on diverse network environments and protect the privacy of individual datasets.
\end{abstract}

\section{Introduction}

Cybersecurity plays a very crucial role in our lives, including social, economic, and political systems. Moreover, with the rise and emergence of the internet, Cybersecurity will become more important than ever as there are various instances of network attacks occurring regularly. Cyberattacks cause organizations great financial, and reputation harm. To name one example: Maersk, one of the world’s largest shipping companies, suffered massively from 2017’s NotPetya malware outbreak. Lack of an efficient Network Intrusion Detection System (NIDS) has been a major factor contributing to the Cyberattacks happening.
\par
There has been ample amount of research recently in the field of NIDS using various supervised machine learning and deep learning algorithms \cite{b3,b4}. With the advancement of AI, many of the data privacy challenges have also been cited, especially when working on real-world tasks with the existing data \cite{b5,b6}. Being able to re-identify information using large datasets, lack of transparency in the use of data, regulations by countries, and unions have been some major factors in intensifying this problem. Thus, there is an inherent risk and higher probability of sensitive information getting exposed and shared. 
\par
Federated Learning (FL) was first proposed by Google to solve problems of data security and privacy in the field of Machine Learning \cite{b8}. Recent years have seen a widespread application of FL methods in various areas \cite{b9,b10}. It showed that by using FL, workers could share intelligence on a machine learning task without disclosing their private data. However, FL has mostly stressed on combining parameters/updates from individual models trained on different samples of data, but FL builds a single global model group to which the individual workers share their updates. However, there might be workers' who wish to build a global model group based on the similarity. Consider the task of intrusion detection, where different networks do not necessarily possess uniform environments. The best system should be able to adapt itself and create global model groups based on the performance and network environment similarity. However, if these segmented global models are not created, quality collaborative learning from such decentralized diverse network environments is a challenging task.
\par
In this paper, we propose the use of an adaptive learning approach of Segmented-Federated Learning (Segmented-FL) for better learning from decentralized diverse network environments. We employ periodic local model evaluation and segmentation for adaptive model training. Unlike FL, Segmented-FL has a feature that each segmented group of workers is arranged with a particular global model for adaptive learning. Segmented-FL is used for parameter sharing among the workers as well as automatic segmentation of workers for adapting to diverse network environments. For each round, selected workers perform model training on their private data and upload their model parameters to their particular global model group where the parameters from various local models are aggregated to update the global model. The parameter aggregation is based on the weighted aggregation of local model parameters according to number of samples each local worker possesses as well as averaging (mean) of the other global model parameters. Periodic evaluation of local workers' model is carried out to validate and divide the workers into different groups for adaptive learning based on recent performance. We prove the effectiveness of our architecture when working with artificial neural networks as our local workers' model. The proposed architecture is found to be superior in handling the diverse network environments while protecting the data privacy and also provides optimum results on publicly available datasets \cite{b28,b29}. Our main contributions in this paper could be listed as: 
\par
\begin{enumerate}
  \item We have made effective use of segmentation to make our system more adaptive and robust to diverse environments.  
  \item We made apt use of the number of samples based weighted aggregation of local model parameters to augment the performance of our system.
  \item We have been able to achieve better performance through the Segmented-FL system, as compared to the conventional FL and centralized approaches. 
\end{enumerate}
\par
The rest of the paper is structured as follows: Section 2 talks about related work in this area while the proposed methodology is explained in section 3. Dataset description and the results obtained are discussed in section 4. Our analysis of the work conducted is presented in section 5 while the paper is concluded in section 6. 

\section{BACKGROUND AND RELATED WORK}
\par
Network intrusion detection has been a much-discussed topic in the industry but the number of publicly available and efficient works are rather limited. Large number of approaches have been tried in the past to solve issues faced in the area of NIDS chiefly due to the rise in the use of machine learning algorithms. One of the earlier and commonly used methods was based on expert knowledge of known malicious patterns and network behavior in a network \cite{b11,b12}. In \cite{b13,b14}, network-based activity study is used to determine if a node is compromised, such as traffic or frequency analysis, and deep packet inspection. However, these approaches seem to be insufficient due to issues of privacy and adaptivity. Lately, various machine learning and neural networks have been implemented for better performance of NIDS systems, such as artificial neural networks (ANN) and support vector machine (SVM) \cite{b16}. \cite{b17} proposed a hybrid deep belief network and SVM for intrusion detection on the NSL-KDD dataset. \cite{b19} proposed a combined approach of employing Boltzmann machine to extract high-level feature representation of network traffic and classifying these features with SVM. \cite{b20} used two-dimensional binary program features based on deep neural network for malware detection. \cite{b21} used unsupervised learning model, named as auto encoder to extract latent features of network traffic for malware detection.
\par
The majority of the work is done in limited environments. Recently, \cite{b22} proposed a federated learning (FL) based cybertattack detection model approach to enhance the performance, protect the privacy of workers, and reduce the traffic load. \cite{b23} also implemented FL to overcome the problem of data insufficiency and data security, where multiple participants collaboratively train a global model. \cite{b24} proposed Cartel, a system to learn collaboratively in edge clouds. A notable observation is that all of these approaches have not taken into consideration the network diversity issue and have proposed single global model based approaches. \cite{b25} presented a Mobility-aware Proactive edge Caching scheme based on Federated learning (MPCF) for predicting content popularity with the private training data distributed on local vehicles. This scheme could adapt to various mobility patterns and references of vehicles and protect users’ privacy. However, unlike former research, we propose a Segmented-FL approach for adaptive intrusion detection in distributed networks.
\section{METHODOLOGY}

\subsection{Data preprocessing and Local model training}
\par
\subsubsection{Data preprocessing}
The data is stripped of superfluous features which may create biases and will not be useful to detect the attacks accurately and reliably. Further, a label encoder is applied to convert the categorical features into numerical form. The retained features dataset is normalized by using MinMax Scaler, which computes the minimum and maximum values of the feature and scales the data to 0-1 range. We perform normalization on our data as input variables affect algorithms that fit a model, which uses a weighted sum of input variables.
\par
On analysis, it was found that the data is imbalanced, and to reduce the variance in the training data we undersample the data using Near Miss 3 \cite{b26}. It operates by only selecting the closest samples from the majority class for each minority class. In 2011, \cite{b27} presented an oversampling approach known as Synthetic Minority Oversampling Technique (SMOTE). However, oversampling too many samples may lead to the creation of inaccurate data samples. Therefore, to overcome the problem of class imbalance we used undersampling, to obtain an accurate and balanced data distribution. Initially, the data had a ratio close to 17:1.2:1 for normal to the attacker to victim class. By selecting only the nearest samples from the normal class, we were able to form a data distribution with a ratio of 2:1.2:1.

\subsubsection{Local Model Training}
We propose the use of a light-weight 3 layer neural network, a common architecture, which will be the primary component of all the local workers. The common model architecture is as shown in fig 1. At local workers, the private data present is preprocessed using the method explained above seamlessly. Further, the workers will train their model on the preprocessed data they possess, whose weights will contribute to the development of a shared global model. As the task in hand is multi-class classification, local models use categorical cross-entropy loss function. 

\begin{figure}[htbp]
\centerline{\includegraphics[width=4cm,height=4cm]{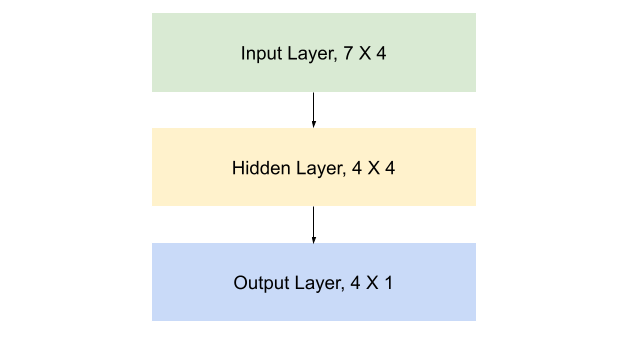}}
\caption{Block diagram of common architecture model}
\label{fig 1}
\end{figure}

\subsection{Federated Learning (FL)}
\par
Federated learning \cite{b8} enables us to learn collaboratively by sharing intelligence on model training with others without sharing their private data. In traditional machine learning algorithms, intrusion detection in networks can expose private information to the outside world while analyzing the data at the central server. The implementation of FL in network intrusion allows us to learn collaboratively by sharing local model training results without accessing any users' private data. The structure and architecture of FL is illustrated in fig 2.
\begin{figure}[htbp]
\centerline{\includegraphics[width=6.5cm,height=5cm]{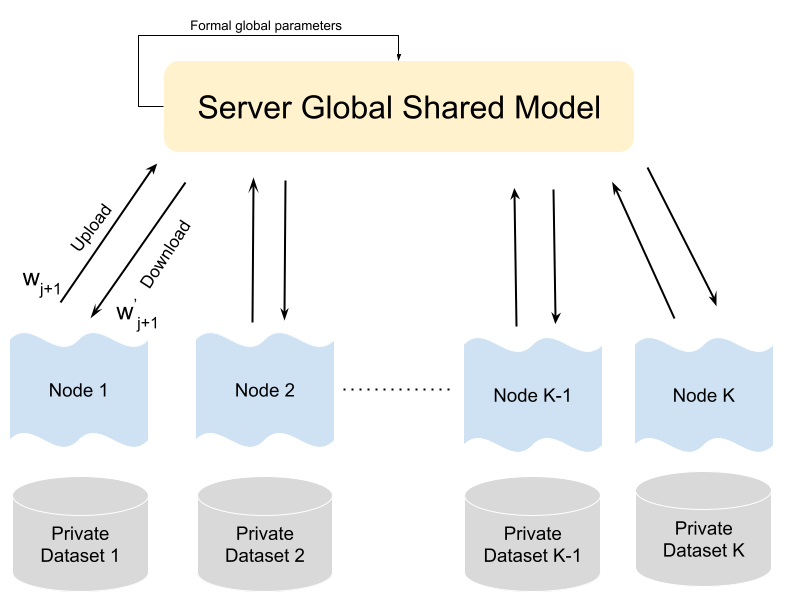}}
\caption{Block diagram of the FL framework. $w_{j+1}$ represents the current round network parameters that upload to the server, $w^{’}_{j+1}$ represents the parameters that are aggregated by server.}
\label{fig 2}
\end{figure}
\par
In this figure, initially, the worker participants download the latest global shared model and update their local models. Then, every worker trains their model by using the local private training data. After training, these trained models' parameters will be uploaded to the central global model for aggregation. Further, based on the aggregated local parameters and former shared global model weights, the global model is updated for the next iteration of collaborative learning. 

\subsection{Segmented-FL Model Architecture}
\par
We propose the use of a novel architecture, Segmented-FL as the principal learning mechanism. By implementing the Segmented-FL system, the worker/network diversity problem is resolved. Unlike FL, Segmented-FL enables us to adapt to various network environments, as well as safeguard the data privacy of the worker nodes by implementing multiple global models. The structure of Segmented-FL is illustrated in fig 3. The working of components is as follows: 
\begin{figure}[htbp]
\centerline{\includegraphics[width=8cm,height=5.5cm]{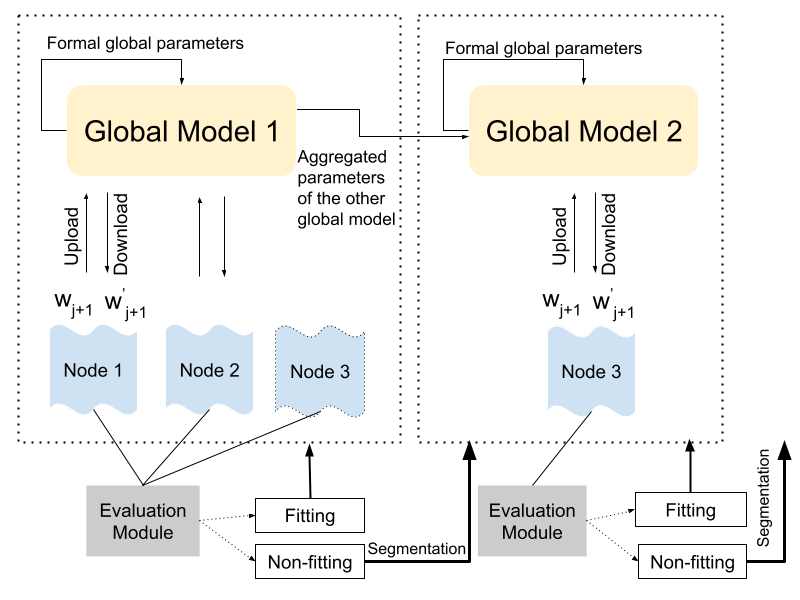}}
\caption{Block diagram of the Segmented-FL framework. }
\label{fig 3}
\end{figure}
\subsubsection{Evaluation Module}
\par
According to fig 3, periodic evaluation is carried out. We have used the F1 score as the metric for periodic evaluation. The periodic evaluation is vital for quality verification of local model training, based on the evaluation module defined in \eqref{eq:label1}. Average validation result is used to evaluate the performance of each worker's local model for training and sharing with the current global model. If the worker's model perfectly fits in the current global model group based on the evaluation module, it remains in the global model group, but if the worker's model does not fit into the current global model group, it is moved to another group. If there is no group created or the worker's model does not fit into any group, a new group is initialized based on the average aggregation result of the workers' who do not fit into any global model group. A flexible threshold is applied, and an evaluation result below the threshold indicates that the worker does not fit into the group. A high threshold brings more workers to segmentation, whereas a lower threshold brings fewer workers for segmentation. The workers moved into the new global model group, conduct the next round's collaborative learning with the initialized global model. We can also set the maximum global model number to a specific low number for simplicity. 
\newline
\begin{equation}
\!
\begin{aligned}
d_{i}=&C_{i}-\frac {\sum ^{n}_{i=1}{C_{i}}}{n} \\ c_{i}=&\frac {1}{1+c^{-d_{i}}} \\ threshold=&0.5- h_{f} \times 0.01
\end{aligned}\label{eq:label1}
\end{equation}
where \\
$n$ is the number of participants, and 
$C_{i}$ represent the average validation result of worker i in the recent rounds, \\
$d_{i}$ is the difference between the worker i's performance and the average performance of all group workers, \\
$c_{i}$ is the output of the evaluation module using the Sigmoid function to convert $d_i$ into the interval of (0, 1), and \\
$h_f$ is the segmentation fineness to adaptively adjust the threshold 

\par
After experimental results, we found out threshold $(h_f) = 7$ with evaluation frequency = 3 to be ideal values in our case.

\subsubsection{Global model aggregation and parameters updation}
\par
 At each communication round j = 1, 2, ..., workers in the current global model group update their models parallelly and send the updated model to the central server for aggregation. Federated Averaging (FedAvg) algorithm is the generic algorithm, which aggregates the workers parameters by averaging (mean). However, averaging the parameters of local workers based on mean affects the performance of the global model where the data is distributed heterogeneously and local workers with very little training examples($n_k$) contribute equally as other local workers. To overcome the inefficiencies caused by the FedAvg algorithm in heterogeneous data distribution, we proposed an aggregating method based on the number of data samples. For the parameter aggregation and updating of the global model, we apply the former global parameters, the weight-aggregated local model parameters, and the average-aggregated parameters of the other global models with different component ratios, defined in \eqref{eq:label3}
 \begin{equation}
\!
\begin{aligned}
   w_{j}=&\alpha \cdot w_{j-\mathrm {1}}+\beta \cdot \frac {\sum ^{k}_{i=1}{n_{i}w_{i}}}{n} \\&+ \gamma \cdot \frac {\sum ^{m}_{i=1} {v_{i}}}{m}\left ({\alpha +\beta +\gamma =1}\right)
\end{aligned}\label{eq:label3}
\end{equation}
where \\
$w_j$ represents the updated global parameters, \\
$w_{j-1}$ represents the former global parameters, \\
$w_i$ represents the local worker i's parameters, \\
$n_i$ represents the number of data samples worker i posses, \\
$n$ represents total data samples present across all workers, \\
$v_i$ represents the global model i's parameters, \\
$k$ is the number of participant workers who conducted local model training in this round, and $m$ is the number of the other global models, \\
$\alpha$, $\beta$, and $\gamma$ represent the ratios of each component respectively, with a sum of 1.

\par
We perform a weighted aggregation of the local models as increasing the weight of strong classifiers enables the formation of a better global shared model. 
\par
\SetAlgoNoLine%
\begin{algorithm}[!h]
 \caption{Segmented-FL algorithm: $w_g=1$  is the initialized global model. $N_g$ is the number of workers related to global model $g$. $N_t$ is the number of workers conducting local model training in a group. $D_t$ is the local dataset of worker t. $G$ represents all global models. $h_j$ is the number of rounds for periodic local model evaluation. $L_g$ is a list including segmentation information of workers. $B$ is the batch size. $E$ is the local training epoch. $\eta$ is the learning rate. $\alpha$, $\beta$ and $\gamma$ are the component ratios for aggregation. $w_{new}$ is the newly initialized global model from the segmentation.}

     \hspace{5pt}initialize $w_{g = 1}$ \newline
 \hspace{5pt}\For{each round j = 1,2,...}{
    \For{each global model g = 1,2,...}{
        $B_t\leftarrow$ (split $N_g$ into batches with a size $N_t$) \\
        $s_t \leftarrow$ (workers conducting model training from $B_t$) \\
        $s_r \leftarrow$ (the other workers not conducting model training) \\
        \For{each worker $t \in s_t $ \textbf{in parallel}}{
            Execute($w_t$, $D_t$) \\
        }
        \For{each worker $r \in s_r $ \textbf{in parallel }}{
                $w_r \leftarrow w_g $ \\
                $w_g \leftarrow $ Aggregate($w_t, t \in s_t; w_g, g \in G $ \\
        }
        \If{$j \% h_j == 0$} {
            \For{each worker $k \in s_g$}{
                $E_k \leftarrow $ avg(validation results of $k$ in recent $R_e$ rounds) \\
                $L_g \leftarrow $ Segment($E_k, k \in s_g; L_g$) \\
            }
        }
    }
 }
\end{algorithm}
\section{DATASET DESCRIPTION AND RESULTS}
\par
The proposed architecture is evaluated on the combination of publicly available standard datasets for network intrusion detection, Coburg intrusion detection dataset (CIDDS-001 and CIDDS-002) \cite{b28,b29}.
\subsection{Dataset Description}
\par
The datasets are generated by emulating a small business environment using  OpenStack. The datasets are labeled as flow-based datasets and contain unidirectional NetFlow data. The dataset consists of 10 features and manually labeled classes and attack types, out of which features such as source IP address, destination IP address, and date were not used. Throughout the study, we focused only on the seven remaining features, and classes as we can see in table 1. We also dropped samples from our dataset, which had classes other than the normal, attacker, and the victim, as the number of data samples having such labels were insignificant in comparison to the other classes samples. Also, we tried using the techniques of undersampling and oversampling on other classes, but the result was not satisfactory. The combined dataset consists of approximately 31 million samples, out of which around 28 million of them are labeled as normal. After dropping other class samples and employing undersampling the majority class, we get 6,472,054 (6.4 million) data samples with a ratio of 2:1.2:1 for normal to attacker to victim classes.

\begin{table}[htbp]
\centering
\caption{Attributes description of Coburg intrusion detection dataset (CIDDS)-001 and CIDDS-002) dataset.}
\label{table 1}
\begin{tabular}{|c|c|} 
\hline
\textbf{Attributes} & \textbf{Description}                                                                   \\ 
\hline
Duration            & Duration of the flow                                                                   \\ 
\hline
Protocol            & \begin{tabular}[c]{@{}c@{}}Transport Protocol \\(e.g. ICMP, TCP, or UDP)\end{tabular}  \\ 
\hline
Source Port         & Source Port                                                                            \\ 
\hline
Destination Port    & Destination Port                                                                       \\ 
\hline
Packets             & Number of transmitted packets                                                          \\ 
\hline
Bytes               & Number of transmitted bytes                                                            \\ 
\hline
Flags               & OR concatenation of all TCP Flags                                                      \\ 
\hline
\textbf{Class}      & \begin{tabular}[c]{@{}c@{}}Class label \\(normal, attacker, or victim)\end{tabular}    \\
\hline
\end{tabular}
\end{table}

\par
We distributed the combined data of CIDDS-001 and CIDDS-002 dataset heterogeneously among workers and applied a train-test split of 90\% and 10\% on each worker, and the technique discussed above to overcome class imbalance is applied. We train our system with 4, and 5 workers, and the data is distributed as shown in table 2. In the case of 4 workers, we distribute the CIDDS-001 among node 1, and node 2 whereas node 3, and 4 contained data majorly from the CIDDS-002. Similarly, for 5 workers, we distribute the CIDDS-001 among node 1, node 2, and node 3, while node 4 and node 5 contain maximum data from the CIDDS-002.

\begin{table}[htbp]
\caption{Data distribution among workers}
\centering
\begin{tabular}{|c|c|c|c|}
\hline
\textbf{No. of workers}         & \textbf{Worker} & \textbf{Training} & \textbf{Testing} \\ \hline
\multirow{4}{*}{\textbf{n = 4}} & Node 1          & 873,727                & 97,081                \\ \cline{2-4} 
                                & Node 2          & 2,038,697              & 226,522               \\ \cline{2-4} 
                                & Node 3          & 1,747,454              & 194,162               \\ \cline{2-4} 
                                & Node 4          & 1,164,970              & 129,441               \\ \hline
\multirow{5}{*}{\textbf{n = 5}} & Node 1          & 698,982                & 77,664                \\ \cline{2-4} 
                                & Node 2          & 1,980,448              & 220,050               \\ \cline{2-4} 
                                & Node 3          & 815,479                & 90,609                \\ \cline{2-4} 
                                & Node 4          & 1,456,212              & 161,802               \\ \cline{2-4} 
                                & Node 5          & 873,727                & 97,081                \\ \hline
\end{tabular}
\label{table 2}
\end{table}



\subsection{Results obtained}
\par
Firstly, different machine learning approaches were applied to the combined CIDDS-001 and CIDDS-002 datasets. The results of each approach are evaluated on the test set on each of the metrics discussed previously.
\par
The results obtained by 4 participants/workers through centralized, federated, and Segmented-FL approaches are shown in table 3. In table 4, we compare results obtained by the respective approaches per label across precision, recall and F1 score. For simplicity, we compare the average of the scores obtained by the workers across the metric.

\begin{table}[!htbp]
\centering
\caption{Comparison across multiple metrics(n = 4)}
\label{table 3}
\begin{tabular}{|c|c|c|c|} 
\hline
\textbf{Approach}                                                                                                    & \textbf{Worker} & \textbf{Accuracy} & \textbf{AUROC}  \\ 
\hline
\multirow{4}{*}{\textbf{Centralized}}                                                                                & Node 1          & 96.89                  & 97.68              \\ 
\cline{2-4}
                                                                                                                     & Node 2          & 97.6                   & 98.4               \\ 
\cline{2-4}
                                                                                                                     & Node 3          & 95.36                  & 96.59               \\ 
\cline{2-4}
                                                                                                                     & Node 4          & 96.04                  & 96.97               \\ 
\hline
\multirow{4}{*}{\begin{tabular}[c]{@{}c@{}}\textbf{Federated}\\\textbf{ Learning}\end{tabular}}                      & Node 1          & 96.63                  & 0.95                 \\ 
\cline{2-4}
                                                                                                                     & Node 2          & 96.92                  & 0.96                 \\ 
\cline{2-4}
                                                                                                                     & Node 3          & 95.49                  & 0.94                 \\ 
\cline{2-4}
                                                                                                                     & Node 4          & 96.16                  & 0.95                 \\ 
\hline
\multirow{4}{*}{\begin{tabular}[c]{@{}c@{}}\textbf{Segmented}\\\textbf{ Federated}\\\textbf{ Learning}\end{tabular}} & Node 1          & \textbf{97.72}         & \textbf{0.97}        \\ 
\cline{2-4}
                                                                                                                     & Node 2          & \textbf{98.58}         & \textbf{0.98}        \\ 
\cline{2-4}
                                                                                                                     & Node 3          & \textbf{96.64}         & \textbf{0.95}        \\ 
\cline{2-4}
                                                                                                                     & Node 4          & \textbf{97.12}         & \textbf{0.96}        \\
\hline
\end{tabular}
\end{table}

\begin{table}[!htbp]
\centering
\caption{Comparison of approaches per label score for n = 4 (FL refers to Federated Learning and Seg-FL refers to Segmented-Federated Learning)}
\label{table 4}
\begin{tabular}{|c|c|c|c|c|} 
\hline
\textbf{Metric}                                                                                & \textbf{Approach} & \textbf{Normal} & \textbf{Attacker} & \textbf{Victim}  \\ 
\hline
\multirow{3}{*}{\begin{tabular}[c]{@{}c@{}}\textbf{Average }\\\textbf{Precision}\end{tabular}} & Centralized       & 0.95            & 0.96              & 0.95             \\ 
\cline{2-5}
                                                                                               & FL                & 0.95            & 0.96              & 0.94            \\ 
\cline{2-5}
                                                                                               & Seg-FL      & \textbf{0.96}   & \textbf{0.97}     & \textbf{0.97}    \\ 
\hline
\multirow{3}{*}{\begin{tabular}[c]{@{}c@{}}\textbf{Average }\\\textbf{Recall}\end{tabular}}    & Centralized       & 0.85            & 0.86              & 0.85             \\ 
\cline{2-5}
                                                                                               & FL                & 0.85            & 0.86              & 0.84             \\ 
\cline{2-5}
                                                                                               & Seg-FL      & \textbf{0.87}   & \textbf{0.88}     & \textbf{0.87}    \\ 
\hline
\multirow{3}{*}{\begin{tabular}[c]{@{}c@{}}\textbf{Average }\\\textbf{F1 Score}\end{tabular}}  & Centralized       & 0.89            & 0.91               & 0.89             \\ 
\cline{2-5}
                                                                                               & FL                & 0.88          & 0.90              & 0.88            \\ 
\cline{2-5}
                                                                                               & Seg-FL      & \textbf{0.91}   & \textbf{0.92}     & \textbf{0.91}    \\
\hline
\end{tabular}
\end{table}

\begin{figure}[!htbp]
\centerline{\includegraphics[width=6cm,height=5cm]{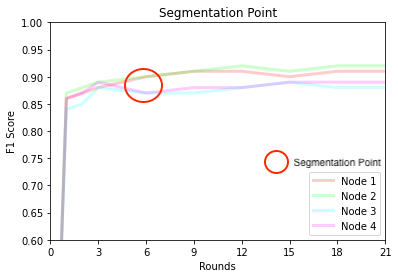}}
\caption{Segmentation Points for n = 4 with threshold ($h_f = 7$)}
\label{fig 4}
\end{figure}

\par
 It can be seen that the Segmented-FL algorithm outweighs the centralized approach models and FL algorithm in all the metrics. The drop in the performance of the FL algorithm could be due to the distribution of the combined dataset, but our Segmented-FL approach has given better results in comparison to other approaches. Further, fig 4 shows that after six rounds of learning, a new global model group is being formed by nodes 3 and 4 as their F1 score was below the threshold of the first global model group in the Segmented-FL approach. In the next tables and figure, we compare our results for five workers and observe the segmentation point.

\begin{table}[!htbp]
\centering
\caption{Comparison across multiple metrics(n = 5)}
\label{table 5}
\begin{tabular}{|c|c|c|c|} 
\hline
\textbf{Approach}                                                                                                    & \textbf{Worker} & \textbf{Accuracy} & \textbf{AUROC}  \\ 
\hline
\multirow{5}{*}{\textbf{Centralized}}                                                                                & Node 1          & 96.47                  & 97.24                \\ 
\cline{2-4}
                                                                                                                     & Node 2          & 97.36                  & 98.19                \\ 
\cline{2-4}
                                                                                                                     & Node 3          & 96.83                  & 97.71                \\ 
\cline{2-4}
                                                                                                                     & Node 4          & 94.69                  & 95.98                \\ 
\cline{2-4}
                                                                                                                     & Node 5          & 93.93                  & 95.07                \\ 
\hline
\multirow{5}{*}{\begin{tabular}[c]{@{}c@{}}\textbf{Federated}\\\textbf{ Learning}\end{tabular}}                      & Node 1          & 96.21                  & 97.35                \\ 
\cline{2-4}
                                                                                                                     & Node 2          & 96.96                  & 97.97                \\ 
\cline{2-4}
                                                                                                                     & Node 3          & 96.53                  & 97.61                \\ 
\cline{2-4}
                                                                                                                     & Node 4          & 94.79                  & 95.88                \\ 
\cline{2-4}
                                                                                                                     & Node 5          & 94.03                  & 95.13                \\ 
\hline
\multirow{5}{*}{\begin{tabular}[c]{@{}c@{}}\textbf{Segmented}\\\textbf{ Federated}\\\textbf{ Learning}\end{tabular}} & Node 1          & \textbf{97.39}         & \textbf{98.41}       \\ 
\cline{2-4}
                                                                                                                     & Node 2          & \textbf{98.42}         & \textbf{99.25}       \\ 
\cline{2-4}
                                                                                                                     & Node 3          & \textbf{97.91}         & \textbf{98.94}       \\ 
\cline{2-4}
                                                                                                                     & Node 4          & \textbf{96.85}         & \textbf{97.73}       \\ 
\cline{2-4}
                                                                                                                     & Node 5          & \textbf{96.13}         & \textbf{97.31}       \\
\hline
\end{tabular}
\end{table}

\begin{table}[!htbp]
\centering
\caption{Comparison of approaches per label score for n = 5 (FL refers to Federated Learning and Seg-FL refers to Segmented-Federated Learning)}
\label{table 6}
\begin{tabular}{|c|c|c|c|c|} 
\hline
\textbf{Metric}                                                                                & \textbf{Approach} & \textbf{Normal} & \textbf{Attacker} & \textbf{Victim}  \\ 
\hline
\multirow{3}{*}{\begin{tabular}[c]{@{}c@{}}\textbf{Average }\\\textbf{Precision}\end{tabular}} & Centralized       & 0.94            & 0.96              & 0.95             \\ 
\cline{2-5}
                                                                                               & FL                & 0.94            & 0.95              & 0.95             \\ 
\cline{2-5}
                                                                                               & Seg-FL      & \textbf{0.96}   & \textbf{0.97}     & \textbf{0.97}    \\ 
\hline
\multirow{3}{*}{\begin{tabular}[c]{@{}c@{}}\textbf{Average }\\\textbf{Recall}\end{tabular}}    & Centralized       & 0.84            & 0.85              & 0.85             \\ 
\cline{2-5}
                                                                                               & FL                & 0.84            & 0.85              & 0.84             \\ 
\cline{2-5}
                                                                                               & Seg-FL      & \textbf{0.87}   & \textbf{0.88}     & \textbf{0.87}    \\ 
\hline
\multirow{3}{*}{\begin{tabular}[c]{@{}c@{}}\textbf{Average }\\\textbf{F1 Score}\end{tabular}}  & Centralized       & 0.88            & 0.9               & 0.88             \\ 
\cline{2-5}
                                                                                               & FL                & 0.88            & 0.89               & 0.87             \\ 
\cline{2-5}
                                                                                               & Seg-FL      & \textbf{0.90}   & \textbf{0.92}     & \textbf{0.90}    \\
\hline
\end{tabular}
\end{table}

\begin{figure}[!htbp]
\centerline{\includegraphics[width=6cm,height=5cm]{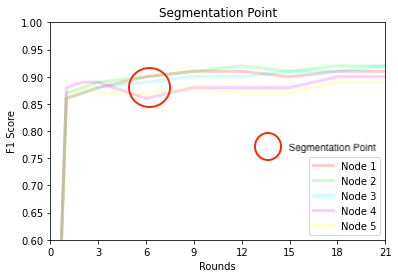}}
\caption{Segmentation Points for n = 5 with threshold ($h_f = 7$)}
\label{fig 5}
\end{figure}

\par
In table 5 and table 6, it can be observed that  even in the case of $n = 5$, the Segmented-FL approach outperforms the FL and Centralized approaches. Also, fig 5 shows that due to variance in F1 score, nodes 4 and 5 (having CIDDS-002 dataset) formed a new global model group after the periodic evaluation at $6^{th}$ round.  Thus, Segmented-FL tends to be more robust to data variance and number of participants/workers, and the result analysis clearly shows the benefit of using Segmented-FL over conventional methods.

\section{ANALYSIS}
Our main observations after performing out the study have been: 
\begin{itemize}
    \item Segmented-FL automatically adapts the architecture based on the periodic evaluation. This feature helps adapt to diverse networks from massively distributed environments.
    \item As compared to the FL approach, Segmented-FL showed more robustness to collaborative learning and had better performance with diverse network traffic.
    \item Due to diverse network environments, there was a drop in the performance of the FL approach for some nodes as compared to the centralized approach. The drop was majorly due to the worker diversity and lower performance from other remaining nodes. This suggests that when workers are diverse, FL can fail to efficiently learn in a collaborative manner.
    \item If all data is to be present at one location, the single centralized model will tend to give better performance than the Segmented-FL and FL approach, but having all data at one location is not possible in every case. Thus, based on the network diversity, Segmented-FL and FL approaches can be implemented to develop a better model. 
\end{itemize}
\section{CONCLUSION}
\par
We proposed the use of the Segmented-FL approach in the area of network intrusion detection. Through this approach, we were able to solve the problem of a network's traffic data not always fitting into the single global model of FL situation. 
Besides, the Segmented-FL approach provided considerably better performance across all the employed metrics as compared to the FL and centralized approaches while preserving the privacy of data. Hence, the improved results across various metrics are a testament to the power which the Segmented-FL approach holds in sensitive and diverse worker training. Future improvements in our work include the use of techniques like differential privacy and a neutral third-party aggregation server to overcome the minor privacy problems federated learning possesses. The mechanism can also be extended further for different dataset structures where a more advanced aggregation mechanism would be required. With the rise in the computing capability of edge devices (like a mobile phone) and strong legislative rules been enforced on data security and privacy, techniques that allow these devices to train their model remotely and collaborate to the global models without needing to share their actual data would be desired. This will lay the foundation for a more secure, ethical, private, and robust data analysis environment.

\nocite{b3}
\nocite{b4}
\nocite{b5}
\nocite{b6}
\nocite{b8}
\nocite{b9}
\nocite{b10}
\nocite{b11}
\nocite{b12}
\nocite{b13}
\nocite{b14}
\nocite{b16}
\nocite{b17}
\nocite{b19}
\nocite{b20}
\nocite{b21}
\nocite{b22}
\nocite{b23}
\nocite{b24}
\nocite{b25}
\nocite{b26}
\nocite{b27}
\nocite{b28}
\nocite{b29}

\bibliographystyle{named}
\bibliography{ijcai21}

\begin{thebibliography}{}

\bibitem[\protect\citeauthoryear{A.~Salama \bgroup \em et al.\egroup
  }{2011}]{b17}
Mostafa A.~Salama, Heba Eid, Rabie Ramadan, Ashraf Darwish, and Aboul~Ella
  Hassanien.
\newblock Hybrid intelligent intrusion detection scheme.
\newblock {\em Advances in Intelligent and Soft Computing, vol. 96, pp.
  295-302}, 96, 01 2011.

\bibitem[\protect\citeauthoryear{Abeshu and Chilamkurti}{2018}]{b22}
Abebe Abeshu and Naveen Chilamkurti.
\newblock Deep learning: The frontier for distributed attack detection in
  fog-to-things computing.
\newblock {\em IEEE Communications Magazine}, 56(2):169--175, 2018.

\bibitem[\protect\citeauthoryear{Carcano \bgroup \em et al.\egroup
  }{2011}]{b12}
A.~Carcano, A.~Coletta, M.~Guglielmi, M.~Masera, I.~Nai~Fovino, and
  A.~Trombetta.
\newblock A multidimensional critical state analysis for detecting intrusions
  in scada systems.
\newblock {\em IEEE Transactions on Industrial Informatics}, 7(2):179--186,
  2011.

\bibitem[\protect\citeauthoryear{Chawla \bgroup \em et al.\egroup }{2002}]{b27}
Nitesh~V. Chawla, Kevin~W. Bowyer, Lawrence~O. Hall, and W.~Philip Kegelmeyer.
\newblock Smote: Synthetic minority over-sampling technique.
\newblock {\em J. Artif. Int. Res.}, 16(1):321–357, June 2002.

\bibitem[\protect\citeauthoryear{Daga \bgroup \em et al.\egroup }{2019}]{b24}
Harshit Daga, Patrick~K. Nicholson, Ada Gavrilovska, and Diego Lugones.
\newblock Cartel: A system for collaborative transfer learning at the edge.
\newblock In {\em Proceedings of the ACM Symposium on Cloud Computing}, SoCC
  '19, page 25–37, New York, NY, USA, 2019. Association for Computing
  Machinery.

\bibitem[\protect\citeauthoryear{Jagannathan and Wright}{2005}]{b5}
Geetha Jagannathan and Rebecca~N. Wright.
\newblock Privacy-preserving distributed k-means clustering over arbitrarily
  partitioned data.
\newblock In {\em Proceedings of the Eleventh ACM SIGKDD International
  Conference on Knowledge Discovery in Data Mining}, KDD '05, page 593–599,
  New York, NY, USA, 2005. Association for Computing Machinery.

\bibitem[\protect\citeauthoryear{McMahan \bgroup \em et al.\egroup }{2016}]{b8}
H.~B. McMahan, Eider Moore, D.~Ramage, and B.~A.~Y. Arcas.
\newblock Federated learning of deep networks using model averaging.
\newblock {\em ArXiv}, abs/1602.05629, 2016.

\bibitem[\protect\citeauthoryear{Mohassel and Zhang}{2017}]{b6}
Payman Mohassel and Yupeng Zhang.
\newblock Secureml: A system for scalable privacy-preserving machine learning.
\newblock In {\em 2017 IEEE Symposium on Security and Privacy (SP)}, pages
  19--38, 2017.

\bibitem[\protect\citeauthoryear{Obeidat \bgroup \em et al.\egroup }{2019}]{b4}
Ibrahim Obeidat, Nabhan Hamadneh, Mouhammd Alkasassbeh, Mohammad Almseidin, and
  Mazen Alzubi.
\newblock Intensive pre-processing of kdd cup 99 for network intrusion
  classification using machine learning techniques.
\newblock {\em International Journal of Interactive Mobile Technologies
  (iJIM)}, 13:70, 01 2019.

\bibitem[\protect\citeauthoryear{Omrani \bgroup \em et al.\egroup }{2017}]{b16}
Takwa Omrani, Adel Dallali, Belgacem~Chibani Rhaimi, and Jaouhar Fattahi.
\newblock Fusion of ann and svm classifiers for network attack detection.
\newblock In {\em 2017 18th International Conference on Sciences and Techniques
  of Automatic Control and Computer Engineering (STA)}, pages 374--377, 2017.

\bibitem[\protect\citeauthoryear{Premaratne \bgroup \em et al.\egroup
  }{2010}]{b11}
Upeka~Kanchana Premaratne, Jagath Samarabandu, Tarlochan~S. Sidhu, Robert
  Beresh, and Jian-Cheng Tan.
\newblock An intrusion detection system for iec61850 automated substations.
\newblock {\em IEEE Transactions on Power Delivery}, 25(4):2376--2383, 2010.

\bibitem[\protect\citeauthoryear{Ring \bgroup \em et al.\egroup }{2017a}]{b28}
Markus Ring, Sarah Wunderlich, Dominik Grüdl, Dieter Landes, and Andreas
  Hotho.
\newblock Creation of flow-based data sets for intrusion detection.
\newblock {\em Journal of Information Warfare}, 16:40--53, 2017.

\bibitem[\protect\citeauthoryear{Ring \bgroup \em et al.\egroup }{2017b}]{b29}
Markus Ring, Sarah Wunderlich, Dominik Grüdl, Dieter Landes, and Andreas
  Hotho.
\newblock Flow-based benchmark data sets for intrusion detection.
\newblock In {\em Proceedings of the 16th European Conference on Cyber Warfare
  and Security (ECCWS)}, pages 361--369. ACPI, 2017.

\bibitem[\protect\citeauthoryear{Saxe and Berlin}{2015}]{b20}
Joshua Saxe and Konstantin Berlin.
\newblock Deep neural network based malware detection using two dimensional
  binary program features.
\newblock pages 11--20, 10 2015.

\bibitem[\protect\citeauthoryear{Shin \bgroup \em et al.\egroup }{2010}]{b13}
Sooyeon Shin, Taekyoung Kwon, Gil-Yong Jo, Youngman Park, and Haekyu Rhy.
\newblock An experimental study of hierarchical intrusion detection for
  wireless industrial sensor networks.
\newblock {\em IEEE Transactions on Industrial Informatics}, 6(4):744--757,
  2010.

\bibitem[\protect\citeauthoryear{Shingi}{2020}]{b9}
Geet Shingi.
\newblock A federated learning based approach for loan defaults prediction.
\newblock pages 362--368, 11 2020.

\bibitem[\protect\citeauthoryear{Smith \bgroup \em et al.\egroup }{2017}]{b10}
Virginia Smith, Chao-Kai Chiang, Maziar Sanjabi, and Ameet Talwalkar.
\newblock Federated multi-task learning.
\newblock 05 2017.

\bibitem[\protect\citeauthoryear{Tsang and Kwong}{2006}]{b14}
Chi-Ho Tsang and Sam Kwong.
\newblock Multi-agent intrusion detection system in industrial network using
  ant colony clustering approach and unsupervised feature extraction.
\newblock pages 51 -- 56, 01 2006.

\bibitem[\protect\citeauthoryear{Vigneswaran \bgroup \em et al.\egroup
  }{2018}]{b3}
Rahul~K. Vigneswaran, R.~Vinayakumar, K.P. Soman, and Prabaharan
  Poornachandran.
\newblock Evaluating shallow and deep neural networks for network intrusion
  detection systems in cyber security.
\newblock In {\em 2018 9th International Conference on Computing, Communication
  and Networking Technologies (ICCCNT)}, pages 1--6, 2018.

\bibitem[\protect\citeauthoryear{Yang \bgroup \em et al.\egroup }{2017}]{b19}
Jun Yang, Jiangdong Deng, Shujuan Li, and Yongle Hao.
\newblock Improved traffic detection with support vector machine based on
  restricted boltzmann machine.
\newblock {\em Soft Computing}, 21, 06 2017.

\bibitem[\protect\citeauthoryear{Yousefi-Azar \bgroup \em et al.\egroup
  }{2017}]{b21}
Mahmood Yousefi-Azar, Vijay Varadharajan, Len Hamey, and Uday Tupakula.
\newblock Autoencoder-based feature learning for cyber security applications.
\newblock In {\em 2017 International Joint Conference on Neural Networks
  (IJCNN)}, pages 3854--3861, 2017.

\bibitem[\protect\citeauthoryear{Yu \bgroup \em et al.\egroup }{2020}]{b25}
Zhengxin Yu, J.~Hu, G.~Min, Zhiwei Zhao, Wang Miao, and M.~Hossain.
\newblock Mobility-aware proactive edge caching for connected vehicles using
  federated learning.
\newblock {\em IEEE Transactions on Intelligent Transportation Systems}, pages
  1--11, 2020.

\bibitem[\protect\citeauthoryear{Zhang and Mani}{2003}]{b26}
J.~Zhang and I.~Mani.
\newblock {KNN Approach to Unbalanced Data Distributions: A Case Study
  Involving Information Extraction}.
\newblock In {\em {Proceedings of the ICML'2003 Workshop on Learning from
  Imbalanced Datasets}}, 2003.

\bibitem[\protect\citeauthoryear{Zhao \bgroup \em et al.\egroup }{2019}]{b23}
Ying Zhao, Junjun Chen, Di~Wu, Jian Teng, and Shui Yu.
\newblock Multi-task network anomaly detection using federated learning.
\newblock SoICT 2019, page 273–279, New York, NY, USA, 2019. Association for
  Computing Machinery.

\end{thebibliography}

\end{document}